\documentstyle[subeqn,12pt,setspace,fullpage]{article}

\newcommand{\be}{\begin{equation}}
\newcommand{\ee}{\end{equation}}
\newcommand{\beq}{\begin{eqnarray}}
\newcommand{\eeq}{\end{eqnarray}}
\advance\voffset by -1truecm

\begin{document}

\onehalfspacing
\begin{center}
{\Large{On the uncertainty principle for proper time and mass}} \\
\vskip 2em
{\large{\sf B. RAM}} \\ [2mm]
Physics Department, New Mexico State University \\
Las Cruces, NM 88003, USA$^\dagger$\\
and\\
Umrao Institute of Fundamental Research\\
A2/214 Janak Puri, New Delhi, 110 058, India
\end{center}
\vskip 4em

\doublespacing 

\begin{abstract}
In [J. Math. Phys. {\bf 40}, 1237 (1999)] Kudaka and Matsumoto derive
the uncertainty relation $c^2 \Delta m \Delta \tau \geq
\hbar/2$ between the rest mass $m$ and the proper time $\tau$,
by considering the Lagrangian $M(\dot \tau - c^{-1} \sqrt{-g_{\mu
\nu}~\dot x^\mu ~\dot x^\nu}) ~+~ e A_\mu (x) \dot x^\mu$.  In this note
we give an alternative derivation based on a special case of the
time-like geodesic equation obtained using the general relativistic
Lagrangian $g_{\mu \nu} \frac{dx^\mu}{d \tau}\frac{dx^\nu}{d \tau}$.
\end{abstract}

\vskip 4em
\hrule width 3cm
\medskip
$^\dagger$ correspondence address
\newpage

In an interesting paper [1] published in this journal, 
Kudaka and Matsumoto derive the
uncertainty relation 
\be
c^2 \Delta m ~\Delta \tau \geq \frac{\hbar}{2}
\ee
between the rest mass $m$ and the proper time $\tau$, by considering
a Lagrangian in which the proper time is included as a dynamic
variable like the positions $x^i$. Specifically, they
consider the Lagrangian
\be
L = M(\dot \tau - c^{-1} \sqrt{-g_{\mu \nu}(x)~\dot x^\mu ~\dot x^\nu})
~+~ e ~A_\mu (x) ~\dot x^\mu
\ee
and find, as a result of their analysis, that the energy 
$E=mc^2$ is the generalized momentum conjugate to the proper
time, and that $\tau,~E,~x^i,$ and $p_i$ are canonical variables
of the system. Consequently the corresponding operators
\[
\hat \tau, ~\hat E, ~\hat x^i, ~\hat p_i ~~~~(i=1,2,3),
\]
satisfy the commutation relations
\be
[\hat E, \hat \tau]= [\hat x^i, \hat p_i]=i \hbar, 
\ee
the relation $[\hat E, \hat \tau]=i \hbar$ in (3) giving the
uncertainty relation (1).

In the present note we give an alternative derivation of Eq. (1).

It is well-known [2] that in general relativity the equations
governing the time-like geodesics in a space-time with the line
element
\be
ds^2 = g_{\mu \nu} ~dx^\mu ~dx^\nu 
\ee
can be derived from the Lagrangian 
\be
2 {\mathcal L} = g_{\mu \nu} ~\frac{dx^\mu}{d\tau}~\frac{dx^\nu}
{d \tau},
\ee
where $\tau$ is the proper time. For the Schwarzschild space-time, the
Lagrangian is (from now on we use units in which $c=\hbar=G=1$)
\be
{\mathcal L} = \frac{1}{2} \left[ \left(1-\frac{2m}{r} \right)
\dot t ^2 - \frac{\dot r ^2}
{1-\frac{2m}{r}} - r^2 \dot \theta^2 - (r^2 \sin ^2 \theta) \dot \phi
^2 \right],
\ee
where the dot denotes differentiation with respect to $\tau$.
The corresponding canonical momenta are
\subequations
\begin{eqnarray}
p_t &=& ~~\frac{\partial \mathcal L}{\partial \dot t} = 
\left ( 1-\frac{2m}{r} \right )
\dot t, \\
p_r &=& -\frac{\partial \mathcal L}{\partial \dot r} = 
\left (1-\frac{2m}{r} \right )
^{-1} \dot r, \\
p_\phi &=& -\frac{\partial \mathcal L}{\partial \dot \phi} = 
(r^2 \sin^2 \theta) \dot \phi,
\end{eqnarray}
and
\be
p_\theta = -\frac{\partial \mathcal L}{\partial \dot \theta} = 
r^2 \dot \theta.
\ee
\endsubequations
\setcounter{equation}{7}
The resulting Hamiltonian is
\be
H = p_t \dot t -(p_r \dot r + p_\theta \dot \theta + p_\phi
\dot \phi) - \mathcal L = \mathcal L.
\ee
From the equality of the Hamiltonian and the Lagrangian it follows that
\be
H = \mathcal L = \rm{constant}.
\ee
For time-like geodesics, $2 \mathcal L$ has the value +1. Integrating
the equations
\be
\frac{dp_t}{d\tau}=\frac{\partial \mathcal L}{\partial t}=0
~~~~~~\rm{and}~~~~~~
\frac{dp_\phi}{d\tau}=-\frac{\partial \mathcal L}{\partial \phi}=0
\ee
one gets
\be
p_t=\left(1 - \frac{2m}{r} \right) \dot t = \rm{constant} = 
{\mathcal E} ~~(\rm say)
\ee
and
\be
p_\phi = r^2 ~\sin^2 \theta ~\dot \phi = \rm{constant}.
\ee
Moreover, from the equation
\be
\frac{dp_\theta}{d\tau} = \frac{d}{d\tau}(r^2 \dot \theta)
= -\frac{\partial {\mathcal L}}{\partial \theta}
=(r^2 \sin \theta \cos \theta) \dot \phi^2,
\ee
it follows that if we choose $\theta = \pi/2$ when $\dot \theta=0$,
then $\ddot \theta$ will also be zero; and $\theta$ will maintain the
value $\pi/2$. In other words, the geodesic is described in an invariant
plane given by $\theta = \pi/2$. Equation (12) then gives
\be
p_\phi = r^2 \dot \phi = \rm{constant} = L ~~~~(\rm{say}),
\ee
where $L$ denotes the angular momentum about an axis
normal to the invariant plane, say the $x$-$y$ plane. With (11), (14)
and ${\mathcal L}=1/2$, Eq. (6) for the time-like geodesic becomes
\be
\frac{1}{2} \left[ \frac{{\mathcal E}^2}{1-\frac{2m}{r}}
-\frac{\dot r ^2}{1-\frac{2m}{r}}
-\frac {L^2}{r^2} \right] = \frac{1}{2}.
\ee
For the special case when both the constants of integration
${\mathcal E}$ and $L$ are zero, Eq. (15) reduces to
\be
\frac{1}{2}\dot r^2 - \frac{m}{r} = -\frac{1}{2}.
\ee
It is easy to see that Eq. (16) describes the region $r \leq 2m$,
$\dot r$ being zero at $r=2m$.

For the Kerr space-time the Lagrangian in the equatorial plane (for
which $\dot \theta = 0$ and $\theta = $ a constant $=\pi/2$) is [2]
\be
2 {\mathcal L} = \left(1-\frac{2m}{r}\right) \dot t^2
+ \frac{4am}{r} \dot t \dot \phi -\frac{r^2}{\Delta}\dot r^2
-\left[ (r^2+a^2) + \frac{2a^2m}{r} \right ] \dot \phi ^2,
\ee
where $a$ is the angular momentum per unit mass of the inner region
and
\be
\Delta = r^2 -2mr +a^2.
\ee
Following (see pp. 326-328 of ref. [2]) in an analogous manner as in
the Schwarzschild case, one obtains, in terms of the constants of
integration ${\mathcal E}$ and $L$,
\be
\frac{1}{2} \dot r^2- \frac{m}{r}+\frac{1}{2}\left(1-{\mathcal
E}^2\right) \left(1+\frac{a^2}{r^2}\right) + \frac{L^2}{2r^2} -
\frac{m}{r^3} (L-a{\mathcal E})^2=0
\ee
as the equation for the time-like geodesic. For the special case when
both the constants of integration ${\mathcal E}$ and $L$ are zero,
Eq. (19) reduces to
\be
\frac{1}{2} \dot r^2 + \frac{a^2}{2r^2} - \frac{m}{r} = -\frac{1}{2}.
\ee
Note that Eq. (20) reduces to Eq. (16) when $a=0$, signifying that the
Schwarzschild solution is a special case of the Kerr solution. Note
also that Eq. (20) can be rewritten as
\be
\frac{1}{2} \dot {\vec r}~^2 - \frac{m}{r} = -\frac{1}{2},
\ee
$\vec{r}$ being given the meaning of a Euclidean vector. Equation (21)
is simply a generalization of Eq. (16). It describes the region
$r \leq 2 m $, $\dot {\vec r}$ being zero at $r=2m$. With
\[
x_1=r \sin \theta_{in} \cos \phi_{in},~~~~~~
x_2=r \sin \theta_{in} \sin \phi_{in},~~~~~~
x_3=r \cos \theta_{in},
\]
\[
r^2=x_1^2+x_2^2+x_3^2, ~~~~~~{\rm and}~~~~~~ 
v^2=\dot x_1^2+\dot x_2^2+\dot x_3^2
=\dot {\vec r}^2,
\]
Eq. (21) is rewritten as
\be
\frac{1}{2}v^2-\frac{m}{r}=-\frac{1}{2}.
\ee

It is also well-known [3] that the three-dimensional Kepler problem,
Eq. (22), is equivalent to a four-dimensional harmonic
oscillator. Below we briefly sketch how it is so, by use of a matrix
transformation known as the Kustaanheimo-Stiefel (KS) transformation
[4].

Let us define the column matrices
\be
X=\left ( \begin{array}{c}
x_1 \\ x_2 \\ x_3 \\ 0 \end{array} \right )
~~~{\rm and}~~~
S=\left ( \begin{array}{c}
s_1 \\ s_2 \\ s_3 \\ s_4 \end{array} \right ),
\ee
\be
\dot X=\left ( \begin{array}{c}
\dot x_1 \\ \dot x_2 \\ \dot x_3 \\ 0 \end{array} \right )
~~~{\rm and}~~~
\dot S=\left ( \begin{array}{c}
\dot s_1 \\ \dot s_2 \\ \dot s_3 \\ \dot s_4 \end{array} \right ),
\ee
where $x_1,~x_2,~x_3$ and $s_1,~s_2,~s_3,~s_4$ are, respectively the
three-dimensional and four-dimensional Cartesian coordinates, and
$\dot x_1,~\dot x_2,~\dot x_3$ and $\dot s_1,~\dot s_2,~\dot s_3,~\dot
s_4$ are the corresponding velocity components. The KS transformation
which transforms the coordinates is given by
\be
X=AS,
\ee
and that which transforms the velocities (or momenta) is given by
\be
\dot X = \frac{1}{2} s^{-2} A \dot S,
\ee
where $ s^2=s_1^2+s_2^2+s_3^2+s_4^2$,
\be
A= \left ( \begin{array}{crrr}
s_3 & -s_4 &  s_1 & -s_2 \\
s_4 &  s_3 &  s_2 &  s_1 \\
s_1 &  s_2 & -s_3 & -s_4 \\
s_2 & -s_1 & -s_4 &  s_3
\end{array} \right ),
\ee
and
\be
s^{-2} {\tilde A} A = {\bf 1},
\ee
${\tilde A}$ being the transposed matrix and {\bf 1} the unit matrix.
It is to be emphasized that (25) and (26) are {\it independent}
transformations. From Eqs. (25), (26) and (28), one obtains
\be
r^2=x_1^2+x_2^2+x_3^2=s^4
\ee
and
\be
v^2=\dot x_1^2+\dot x_2^2+\dot x_3^2=\frac{1}{4s^2}\dot s^2
=\frac{1}{4s^2}
\left(\dot s_1^2 + \dot s_2^2 + \dot s_3^2 + \dot s_4^2 \right).
\ee
Using (29) and (30) it is straightforward to show that Eq. (21)
transforms into
\be
\frac{1}{2}m_{ho}\dot s^2 + \frac{1}{2}m_{ho}\omega^2 s^2=m,
\ee
with $m_{ho}\equiv$ mass of the four-dimensional harmonic oscillator
=1/4 and $\omega=2$.
 
Note that in Eq. (31) $m$ corresponds to the total classical energy
$(E)$, the proper time $\tau$ to the Newtonian time ($t$), and the
quantum equation that corresponds to (31) is
\be
-\frac{1}{2m_{ho}}\sum^4_{i=1}\left(\frac{\partial^2}{\partial s_i^2}
-m_{ho}^2\omega^2 s^2 \right ) \psi
=(n+1) \psi
\ee
In (32) $(n+1)$, $n=0,1,...$ are the eigenvalues of the operator
$m$. Consequently the Heisenberg uncertainty relation [5] 
$\Delta E \Delta t \geq \hbar / 2$ (inserting $c$ and $\hbar$)
simply translates into
\[
c^2 \Delta m ~\Delta \tau \geq \frac{\hbar}{2}.
\]

For application of the quantum equations that correspond to (16) and
(21) to the black hole physics, the interested reader may refer to
refs. [6-8].

\vskip2cm
\begin{center}
Acknowledgements
\end{center}

The author thanks S.R. Choudhury, N. Mukunda and S.M. Roy for
conversations, R.S. Bhalerao for a critical reading of the manuscript
leading to an improved version and the Tata Institute for a pleasant
stay.

\end{document}